\documentclass[manuscript,screen,authorversion, acmtog]{acmart}

\usepackage{times}
\usepackage{epsfig}
\usepackage{graphicx}

\usepackage{caption}
\usepackage{enumitem}
\usepackage{xspace}
\usepackage{subfigure}
\usepackage{hhline}
\usepackage{xcolor}
\usepackage{balance}
\usepackage{bm}
\usepackage[toc,page]{appendix}
\usepackage{multirow}
\usepackage{mfirstuc}
\usepackage{cleveref}

\newcommand{\video}{{{{video}}}\xspace}

\newcommand{\threeD}{{3D}\xspace}

\newcommand{\mocap}{MoCap\xspace}

\newcommand{\etal}{et al.\xspace}
\newcommand{\ie}{i.e.\xspace}

\newcommand{\wrt}{w.r.t.\xspace}

\usepackage{amsmath,amsfonts,bm}

\def\eqref#1{equation~\ref{#1}}

\def\1{\bm{1}}

\def\rva{{\mathbf{a}}}

\def\rvd{{\mathbf{d}}}

\def\rvg{{\mathbf{g}}}

\def\rvs{{\mathbf{s}}}

\def\rvx{{\mathbf{x}}}

\def\rvmu{{\mathbf{\mu}}}

\DeclareMathAlphabet{\mathsfit}{\encodingdefault}{\sfdefault}{m}{sl}
\SetMathAlphabet{\mathsfit}{bold}{\encodingdefault}{\sfdefault}{bx}{n}

\newcommand{\expec}{\mathbb{E}}

\begin{document}

\title{Synthesizing Physical Character-Scene Interactions}

\author{Mohamed Hassan}
\affiliation{%
 \institution{Electronic Arts}
 \country{USA}}
\email{mohassan@ea.com}
\thanks{The work was done while Mohamed Hassan was an intern at Nvidia.}
\author{Yunrong Guo}
\email{kellyg@nvidia.com}
\affiliation{
  \institution{NVIDIA}
  \country{Canada}
}
\author{Tingwu Wang}
\email{tingwuw@nvidia.com}
\affiliation{
  \institution{NVIDIA}
  \country{Canada}
}
\author{Michael Black}
\email{black@tuebingen.mpg.de}
\affiliation{
  \institution{Max-Planck-Institute for Intelligent Systems}
  \country{Germany}
}
\author{Sanja Fidler}
\email{sfidler@nvidia.com}
\affiliation{
  \institution{University of Toronto}
  \country{Canada}
}
\affiliation{
  \institution{NVIDIA}
  \country{Canada}
}
\author{Xue Bin Peng}
\email{xbpeng@berkeley.edu}
\affiliation{
  \institution{NVIDIA}
  \country{Canada}
  }
  \affiliation{
  \institution{Simon Fraser University}
  \country{Canada}
}

\begin{abstract}
Movement is how people interact with and affect their environment. 
For realistic character animation, it is necessary to synthesize such interactions between virtual characters and their surroundings.
Despite recent progress in character animation using machine learning, most systems focus on controlling an agent's movements in fairly simple and homogeneous environments, with limited interactions with other objects. 
Furthermore, many previous approaches that synthesize human-scene interactions require significant manual labeling of the training data.
In contrast, we present a system that uses adversarial imitation learning and reinforcement learning to train physically-simulated characters that perform scene interaction tasks in a natural and life-like manner. 
Our method learns scene interaction behaviors from large unstructured motion datasets, without manual annotation of the motion data.
These scene interactions are learned using an adversarial discriminator that evaluates the realism of a motion within the context of a scene.
The key novelty involves conditioning both the discriminator and the policy networks on scene context.
We demonstrate the effectiveness of our approach through three challenging scene interaction tasks: carrying, sitting, and lying down, which require coordination of a character's movements in relation to objects in the environment. 
Our policies learn to seamlessly transition between different behaviors like idling, walking, and sitting. %
By randomizing the properties of the objects and their placements during training, our method is able to generalize beyond the objects and scenarios depicted in the training dataset, producing natural character-scene interactions for a wide variety of object shapes and placements.
The approach takes physics-based character motion generation a step closer to broad applicability. Please see our \href{https://youtu.be/q3hyQdaElQQ}{supplementary video} for more results.
\end{abstract}



\begin{teaserfigure}
  \includegraphics[width=\linewidth]{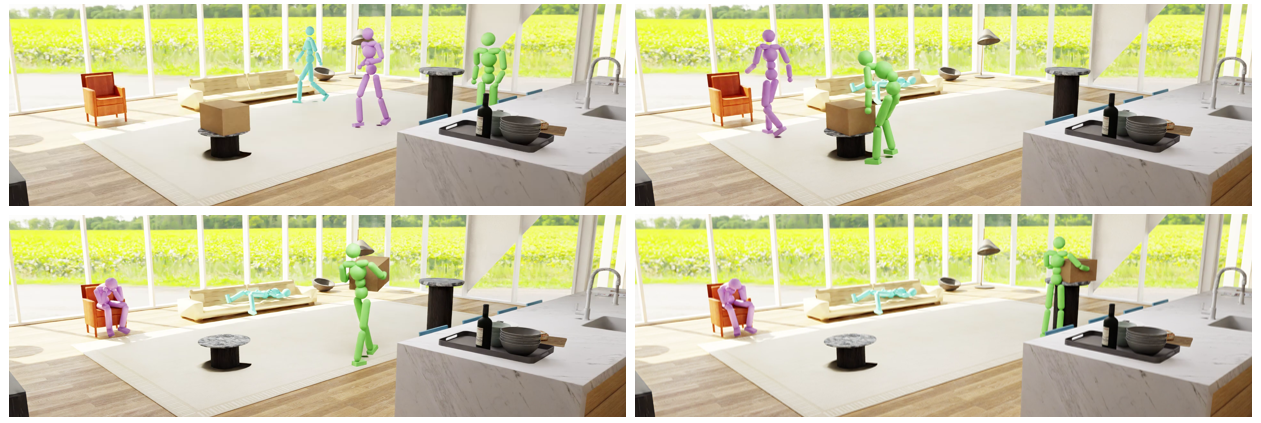}
  \caption{Our framework enables physically simulated characters to perform scene interaction tasks in a natural and life-like manner. We demonstrate the effectiveness of our approach through three challenging scene interaction tasks: carrying, sitting, and lying down, which require coordination of a character's movements in relation to objects in the environment. }
  \label{fig:teaser}
\end{teaserfigure}

\maketitle

\section{Introduction}
Realistically animating virtual characters is a challenging and fundamental problem in computer graphics. 
Most prior work focuses on generating realistic human motions and often overlooks the fact that, when humans move, the movements are often driven by the need to interact with objects in a scene.
When interacting with a scene, characters need to ``perceive" the objects in the environment and adapt their movements by taking into account environmental constraints and affordances.
The objects in the environment can restrict movement, but also afford opportunities for interaction.  Therefore characters need to adapt their movements according to object-specific functionality. 
Lying down on a bunk bed requires different movements than lying down on a sofa.
Similarly, picking up objects of different sizes may require different strategies. %

Existing techniques for synthesizing character-scene interactions tend to be limited in terms of motion quality, generalization, or scalability. Traditional motion blending and editing techniques~\cite{Gleicher_1997, Lee_Shin_199} require significant manual effort to adapt existing motion clips to a new scene. Data-driven kinematic models~\cite{hassan_samp_2021,Holden_2017,mode_adaptive_zhang_2018, nsm_2019} produce high-quality motion when applied in environments similar to those seen during training. However, when applied to new scenarios, such kinematic models struggle to generate realistic behaviors that respect scene constraints. 
Physics-based methods are better able to  synthesize plausible motions in new scenarios by leveraging a physics simulation of a character's movements and interactions within a scene. 
Reinforcement learning (RL) has become one of the most commonly used paradigms for developing control policies for physically-simulated characters. However, it can be notoriously difficult to design RL objectives that lead to high-quality and natural motions~\cite{heess2017emergence}. 
Motion tracking~\cite{peng_2017} can improve motion quality by training control policies to imitate reference motion data. However, it can be difficult to apply tracking-based methods to complex scene-interaction tasks, where a character may need to compose, and transition between, a diverse set of skills in order to effectively interact with its surroundings.

Recently, Adversarial Motion Priors (AMP) \cite{peng_2021} have been proposed as a means of imitating behaviors from large unstructured motion datasets, without requiring any annotation of the motion data or an explicit motion planner. This method leverages an adversarial discriminator to differentiate between motions in the dataset and motions generated by the policy. The policy is trained to satisfy a task reward while also trying to fool the discriminator by producing motions that resemble those shown in the dataset. Crucially, the policy need not explicitly track any particular motion clip, but is instead trained to produce motions that are within the distribution of the dataset. This allows the policy to deviate, interpolate, and transition between different behaviors as needed to adapt to new scenarios. This versatility is crucial for character-scene interaction, which requires fine-grain adjustments to a character's behaviors in order to adapt to different object configurations within a scene.

In this work, we present a framework for training physically simulated characters to perform scene interaction tasks. Our method builds on AMP and extends it to character-scene interaction tasks. %
Unlike the AMP discriminator, which only considers the character's motion, our discriminator jointly examines the character and the object in the scene. This allows our discriminator to evaluate the realism of the character's movements within the context of a scene (e.g., a sitting motion is realistic only when a chair is present).
In addition, given a small dataset of human-object interactions, our policy discovers how to adapt these behaviors to new scenes. For example, from about five minutes of motion capture data of a human carrying a single box, we are able to train a policy to carry hundreds of boxes with different sizes and weights. %
We achieve this by populating our simulated environments with a wide range of object instances and randomizing their configuration and physical properties. By interacting with these rich simulated environments, our policies learn how to realistically interact with a wide range of object instances and environment configurations. We demonstrate the effectiveness of our method with three challenging scene-interaction tasks: sit, lie down, and carry. As we show in our experiments, our policies are able to effectively perform all of these tasks and achieve superior performance compared to prior state-of-the-art kinematic and physics-based methods.

In summary, our main contributions are:
(1) A framework for training physically simulated characters to perform scene interaction tasks without manual annotation.
(2) We leverage a scene-conditioned discriminator that takes into account a character's movements in the context of objects in the environment.
(3) We introduce a randomization approach for physical properties of objects in the scene that enables generalization beyond the objects shown in the demonstration.
While our framework consists of individual components that have been introduced in prior work, the particular choice and combination of these components in the context of physics-based scene interaction tasks is novel, and we demonstrate state-of-the-art results for accomplishing these tasks with physically simulated characters.

\section{Related Work} 
Traditional animation methods generally edit, retarget, or replay motion clips from a database in order to synthesize motions for a given task. The seminal work of \citet{Gleicher_1997} adapts a reference motion to new characters with different morphologies. %
The method is used to adapt scene-interaction motions like box carrying and climbing a ladder. \citet{Lee_Shin_199} introduce an interactive motion editing technique that allows motions to be adapted to new characters and new environments.
Such editing and retargetting methods are limited to new scenarios that are similar to the original source motion clip. 
In the interest of brevity, the following discussion focuses on full body animation. However, there is a long line of related research on dexterous manipulation. See \citet{Sueda_2008, ManipNet, wheatland2015state, Yuting_2012} for more details.

\subsection{Deep Learning Kinematic Methods}
The applicability of deep neural networks (NN) to human motion synthesis has been studied extensively \cite{Fragkiadaki_2015, Habibie2017ARV, holden2016deep, martinez2017human, Taylor_2009}. %
Unlike other regression tasks, classical architectures like CNNs, LSTMs and feed-forward networks perform poorly on motion synthesis. They tend to diverge or converge to a mean pose when generating long sequences.
Thus, several novel architectures have been introduced in the literature to improve the motion quality. 
For instance, instead of directly training a single set of NN parameters,
Phase-Functioned Neural Networks~\cite{Holden_2017} compute the NN parameters at each frame as a function of the phase of a motion. 
This model can generate high-quality motions but is limited to cyclic behaviors that progress according to a well-defined phase variable. 
\citet{nsm_2019} use a phase variable and mixture of experts~\cite{eigen2013learning, Jacobs_MOE} to synthesize object interaction behaviors, such as sitting and carrying. %
SAMP~\cite{hassan_samp_2021} avoids the need for phase labels by training an auto-regressive cVAE~\cite{Kingma2014auto, cvae} using scheduled sampling~\cite{Bengio_2015}.
Instead of manually labelling a single phase for a motion, local motion phase variables can also be automatically computed for each body part using an evolutionary strategy  \cite{Starke_2020}. 
Such data-driven kinematic scene-interaction methods typically require high-quality \threeD human-scene data, which is scarce and difficult to record. Since these methods only learn from demonstrations, their performance degrades when applied to scenarios unlike those in the training dataset~\cite{couch_2022, wang2022towards, zhang2022wanderings, wang2020synthesizing}. %

\subsection{Physics-Based Methods}
Physics-based methods generate motions by leveraging the equations of motion of a system \cite{raibert_19991}.
The physical plausibility of the generated motion is guaranteed, but the resulting behaviors may not be particularly life-like, since simulated character models provide only a coarse approximation of the biomechanical properties of their real-life counterparts. Heuristics, such as symmetry, stability, and power minimization \cite{raibert_19991, wang_2009} can be incorporated into controllers to improve the realism of simulated motions. 
Imitation learning is another popular approach to improve the realism of physically simulated characters. In this approach, a character learns to perform various behaviors by imitating reference motion data \cite{peng_2017}. 
Motion tracking is one of the most commonly used techniques for motion imitation and is effective at reproducing a large array of challenging skills \cite{MocapImitationChentanez2018,DreCon2019,wang2020unicon,ScalableWon2020}.
However, it can be difficult to apply tracking-based methods to solve tasks that require composition of diverse behaviors, since the tracking-objective is typically only applied with respect to one reference motion at a time.
Inspired by Generative Adversarial Imitation Learning (GAIL)~\cite{gail_2016}, \citet{peng_2021} train a motion discriminator on large unstructured datasets and use it as a general motion prior for training control policies.
This technique allows characters to imitate and compose behaviors from large datasets, without requiring any annotation of the motion clips, such as skill or phase labels.
In this work, we leverage an adversarial imitation learning approach, but go beyond prior work to develop control policies for character-scene interaction tasks.

\subsection{Character-Scene Interaction}
Very little work has tackled the problem of synthesizing physical character-scene interactions.
Early work simplifies the object manipulation problem by explicitly attaching an object to the hands of the character \cite{coros_2010, peng_2019,mordatch_2012}, thereby removing the need for the character to grasp and manipulate an object's movements via contact. \citet{liu_2018} use a framework based on trajectory optimization to learn basket-ball dribbling. %
\citet{chao2019learning_to_sit} propose a hierarchical controller to synthesize sitting motions, by dividing the sitting task into sub-tasks and training separate controllers to imitate relevant reference motion clips for each sub-task.
A meta controller is then trained to select which sub-task to execute at each time step.
A similar hierarchical approach is used to train characters to play a simplified version of football \cite{liu2021motor,huang2021tikick}.
\citet{Merel_2020} train a collection of policies, each of which imitates a motion clip depicting a box-carrying or ball-catching task. The different controllers are then distilled into a single latent variable model that can then be used to construct a hierarchical controller for performing more general instances of the tasks. %
In contrast to the prior work, our approach is not hierarchical, generalizes to more objects and scenes, can be trained on large datasets without manual labels, and is easily applicable to multiple tasks.

\section{Method}
To train policies that enable simulated characters to interact with objects in a natural and life-like manner, we build on the Adversarial Motion Priors (AMP) framework \citep{peng_2021}. Our approach consists of two components: a policy and a discriminator as shown in Fig.~\ref{fig:pipline}. The discriminator's role is to differentiate between the behaviors produced by the simulated character and the behaviors depicted in a motion dataset. The role of the policy $\pi$ is to control the movements of the  character in order to maximize the expected accumulative reward $J(\pi)$. The agent's reward $r_t$ at each time step $t$ is specified according to: 
\begin{figure}
	\centering	
	\includegraphics[width=\linewidth]{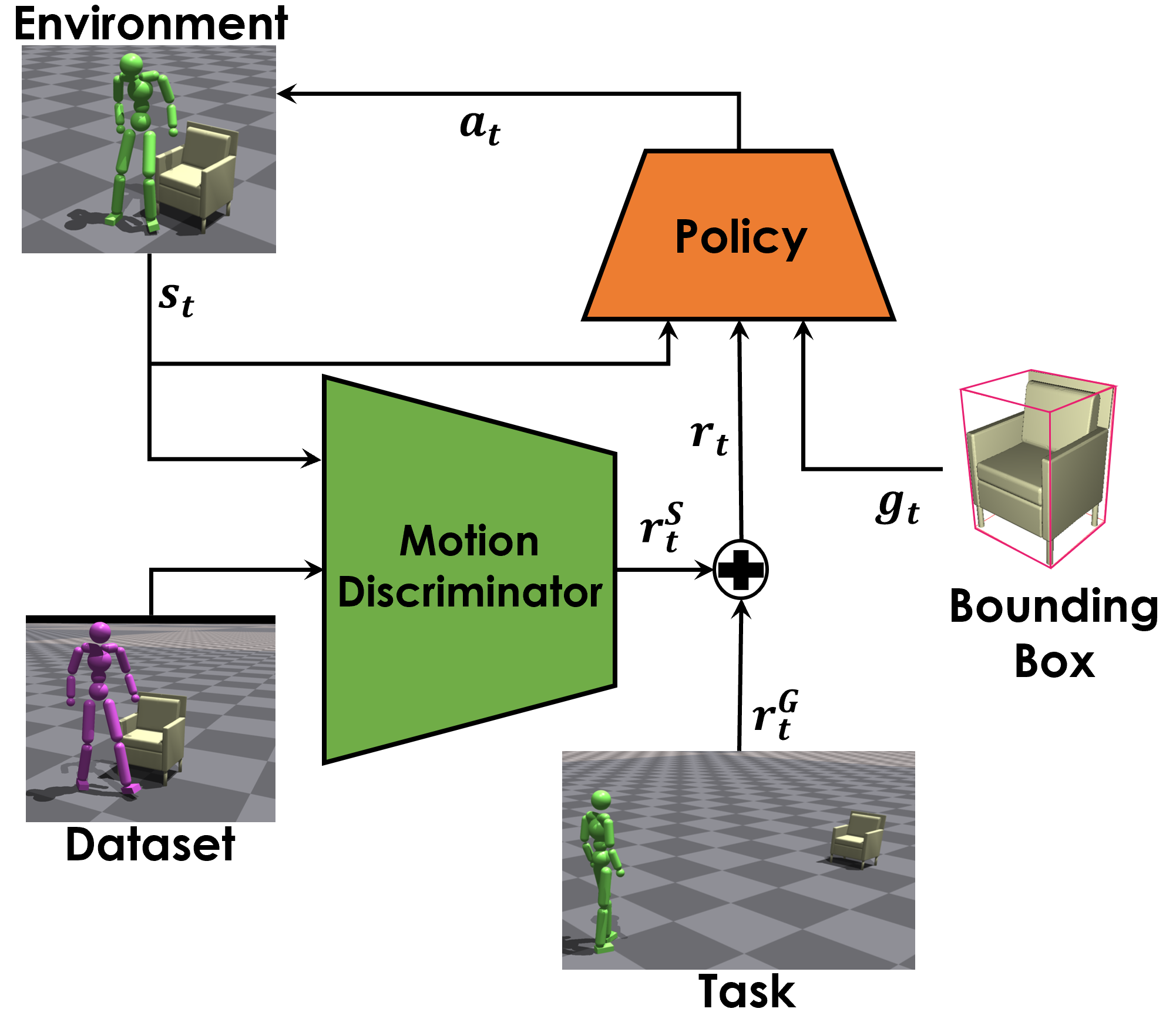}	
	\caption{
	Our framework has two main components: a policy and a discriminator. The discriminator differentiates between the behaviors generated by the policy and the behaviors depicted in a motion dataset. 
	In contrast to prior work, our discriminator receives information pertaining to both the character and the environment.
	Specifically, the policy is trained to control the character movements to achieve a task reward $r^G$ while producing a motion that looks like realistic human behavior within the context of a given scene.
	}
	\label{fig:pipline}
\end{figure}
\begin{equation}
    r_t = w^G r^G(\rvs_t, \rvg_t, \rvs_{t+1}) + w^S r^S(\rvs_t, \rvs_{t+1}).
\label{eqn:reward}
\end{equation}
The task reward $r^G$ encourages the character to satisfy high-level objectives, such as sitting on a chair or moving an object to the desired location. The style reward $r^S$ encourages the character to imitate behaviors from a motion dataset as it performs the desired task.
$\rvs_t \in \mathcal{S}$ is the state at time $t$. $\rva_t \in \mathcal{A}$ are the actions sampled from the policy $\pi$ at time step $t$. $\rvg_t \in \mathcal{G}$ denotes the task-specific goal features at time $t$. $w^G$ and $w^S$ are weights. The policy is  trained to maximize the expected discount return $J(\pi)$,
\begin{equation}
    J(\pi) = \expec_{p(\tau | \pi)} \left[ \sum_{t=0}^{T-1} \gamma^t r_t \right],
\label{eqn:rl_objective}
\end{equation}
where $p(\tau | \pi)$ denotes the likelihood of a trajectory $\tau$ under the policy $\pi$. $T$ is the time horizon, and $\gamma \in [0, 1]$ is a discount factor.

The style reward $r^S$ is modeled using an adversarial discriminator that evaluates the similarity between the motions produced by the physically simulated character and the motions depicted in a dataset of motion clips. The discriminator is trained according to the objective proposed by ~\citet{peng_2021}: 
\begin{align}
    \mathop{\mathrm{arg \ min}}_D \ & -\expec_{d^\mathcal{M}(\rvs, \rvs_{t+1})} \left[ \mathrm{log}\left(D(\rvs, \rvs_{t+1})\right) \right] \\
    & - \expec_{d^\pi({\rvs, \rvs_{t+1}})} \left[\mathrm{log}\left(1 - D(\rvs, \rvs_{t+1})\right) \right] \\
    & + w^\mathrm{gp} \ \expec_{d^\mathcal{M}(\rvs, \rvs_{t+1})} \left[\left| \left| \nabla_\phi D(\phi) \middle|_{\phi = (\rvs, \rvs_{t+1})} \right| \right|^2 \right],
\label{eqn:disc_loss}
\end{align}
where $d^\mathcal{M}(\rvs, \rvs_{t+1})$ and $d^\pi({\rvs, \rvs_{t+1}})$ represent the likelihoods of the state transition from $\rvs$ to $\rvs_{t+1}$ under the dataset distribution $\mathcal{M}$ and the policy $\pi$ respectively. $w^\mathrm{gp}$ is a manually specified coefficient for a  gradient penalty regularizer \citep{mescheder18a}.
The style reward $r^S$ for the policy is then specified according to:
\begin{equation}
    r^S(\rvs_t, \rvs_{t+1}) =  - \mathrm{log}(1 - D(\rvs_t, \rvs_{t+1})) .
\label{eqn:gan_reward}
\end{equation}

\section{State and Action Representation}
The state $\rvs$ is represented by a set of features that describes the configuration of the character's body, as well as the configuration of the objects in the scene relative to the character. These features include:
\begin{itemize}
  \item Root height
  \item Root rotation
  \item Root linear and angular velocity
  \item Local joints rotations
  \item Local joints velocities
  \item Positions of four key joints: right hand, left hand, right foot, and left foot
  \item Object position
  \item Object orientation
\end{itemize}
The height and rotation of the root are recorded in the world coordinate frame while velocities of the root are recorded in the character's local coordinate frame. Rotations are presented using a 6D normal-tangent encoding \citep{peng_2021}. The positions of four key joints, object position, and object orientation are recorded in the character's local coordinate frame. 
A key difference from prior work is the inclusion of object features in the state.
These object features enable the discriminator to not only judge the realism of the motion but also how realistic the motion is \wrt to the object.
Note that the object can move during the action and the agent must react appropriately.
Combined, these features result in a $114$D state space. The actions $\rva$ generated by the policy specify joint target rotations for PD controllers. Each target is represented as an exponential map $\rva \in \mathbb{R}^3$ \citep{ExpMapGrassia1998}, resulting in a  $28$D action space. 

We demonstrate the effectiveness of our framework on three challenging interactive tasks: sit, lie down, and carry. Separate policies are trained for each task.
The style reward $r^S$ is the same for all tasks. Please refer to the supplementary material for a detailed definition of the task-specific reward $r^G$. %

\section{Motion Dataset}
In order to train the character to interact with objects in a life-like manner, we train our method using a motion dataset of human-scene interactions. For the sit and lie down tasks; we use the SAMP dataset \cite{hassan_samp_2021}, which contains $100$ minutes of \mocap clips of sitting and lying down behaviors. Furthermore, the dataset also records the positions and orientations of objects in the scene, along with CAD models for seven different objects. For the carry task; we captured $15$ \mocap clips of a subject carrying a single box. In each clip, the subject walks towards the box, picks it up, and carries it to a target location. The initial and target box locations are varied in each clip. 
In addition to full-body \mocap, the motion of the box is also tracked using optical markers. 

The SAMP dataset provides examples of interactions with only seven objects, similarly our object-carry dataset only contains demonstration of carrying a single box. 
Nonetheless we show that our reinforcement learning framework allows the agent to generalize from these limited demonstrations to interact with a much wider array of objects in a natural manner. This is achieved by exposing the policy to new objects in the training phase. Our policy is trained using multiple environments simulated in parallel in IsaacGym~\cite{makoviychuk2021isaac}. 
We populate each environment with different object instances to encourage our policy to learn how to interact with objects exhibiting natural class variation.
For the sit and lie down tasks we replace the original objects with different objects of the same class from ShapeNet~\cite{shapenet2015}. 
The categories are: regular chairs, armchairs, tables, low stools, high stools, sofas, and beds. 
In total, we used $\sim 350$ unique objects from ShapeNet~\cite{shapenet2015}. To further increase the diversity of the objects, we randomly scale the objects in each training episode by a scale factor between $0.8$ and $1.2$.
For the carry task; the size of the object is randomly scaled by a factor between $0.5$ and $1.5$.

\section{Training}
\label{sec:training}
At the start of each episode, the character and objects are initialized to states sampled randomly from the dataset.
This leads to the character sometimes being initialized far from the target, requiring it to learn to walk towards the target and execute the desired action. 
At other times, it is initialized close to the completion state of the task, \ie sitting on the object or holding a box. In contrast to always initializing the policy to a fixed starting state, this Reference State Initialization approach~\cite{peng_2017} has been shown to significantly speed up training progress and produce more realistic motions.

Since the reference motions depict only a limited set of scenarios, initialization from this alone is not sufficient to cover all possible configurations of the scene. In order to train general policies that are able to execute the desired task from a wide range of initial configurations, we randomize the object position \wrt the character at the beginning of each episode.
The object is placed anywhere between one and ten meters away from the character on the horizontal plane. The object orientation is sampled uniformly between $[0, 2\pi]$.
The episode length is set to $10$ seconds for the sit and lie down tasks, and $15$ seconds for the carry task. 
In addition, we terminate the policy early if any joint, except the feet and hands, is within $20$cm of the ground, or if the box is within $30$cm of the ground. 

The policy $\pi$ is modeled using a neural network that takes as input the current state $\rvs_t$ and goal $\rvg_t$, then predicts the mean $\rvmu(\rvs_t, \rvg_t)$ of a Gaussian action distribution $\pi(\rva_t | \rvs_t, \rvg_t) = \mathcal{N}\left(\rvmu(\rvs_t, \rvg_t), \Sigma \right)$. The covariance matrix $\Sigma$ is manually specified and kept fixed during training. The policy, value function and the discriminator are modeled by separate fully-connected networks with the following dimensions $\left \{1024, 512, 28\right \}$, $\left \{1024, 512, 1\right \}$, $\left \{1024, 512, 1\right \}$ respectively. ReLU activations are used for all hidden units.
We follow the training strategy of \citet{peng_2021} to jointly train the policy and the discriminator.

\begin{figure*}
     \centering
     \subfigure[Sit]{\includegraphics[width=0.33\textwidth]{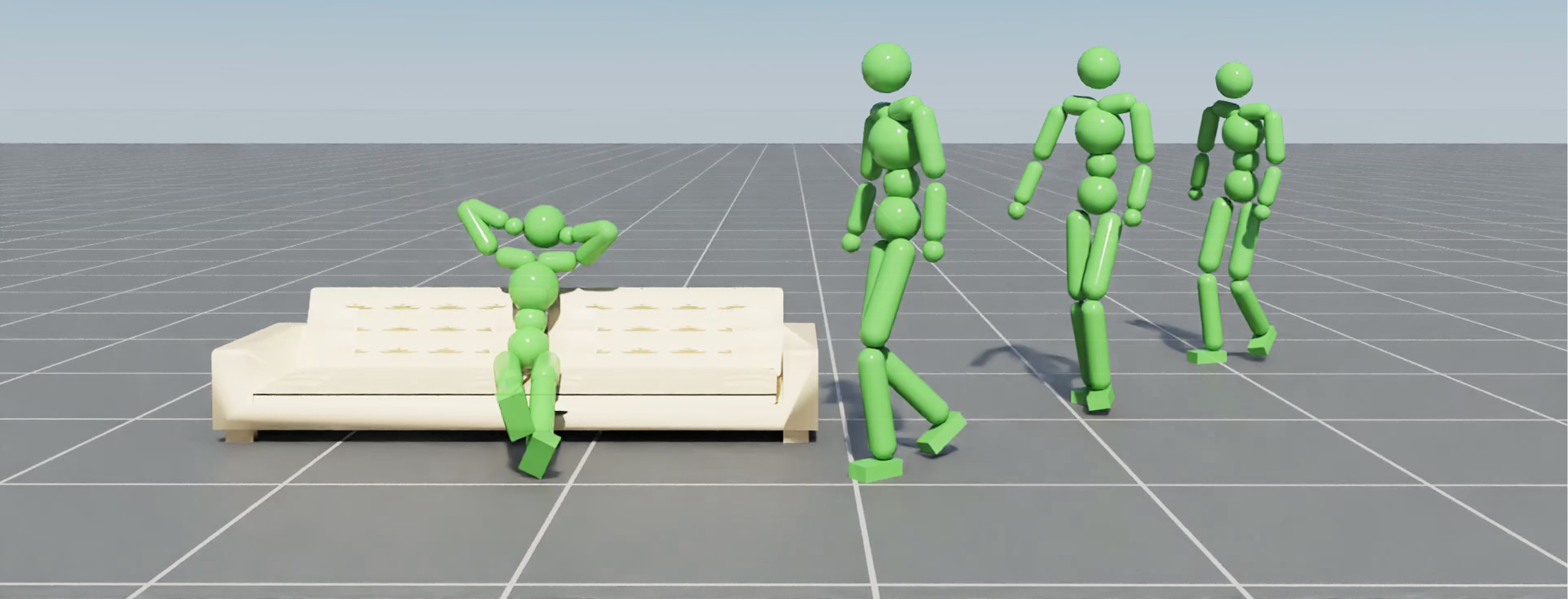}}
     \subfigure[Lie down]{\includegraphics[width=0.33\textwidth]{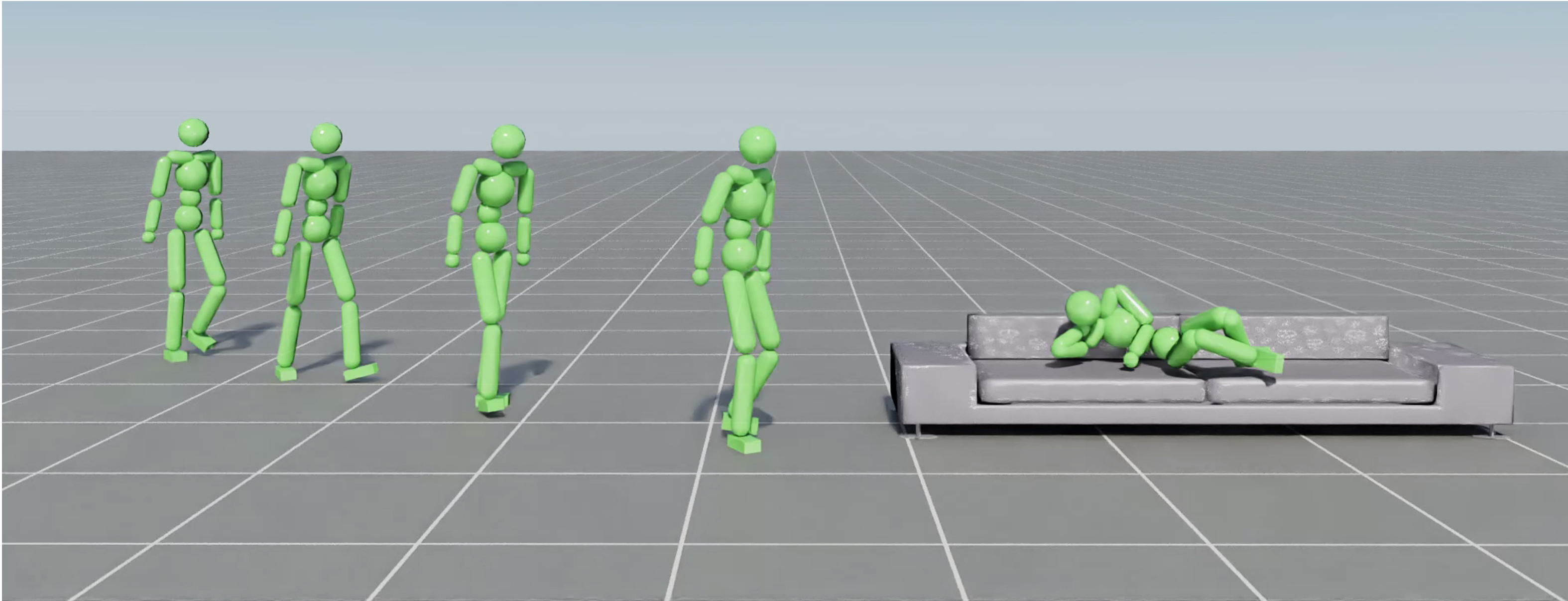}}
      \subfigure[Carry ]{\includegraphics[width=0.33\textwidth]{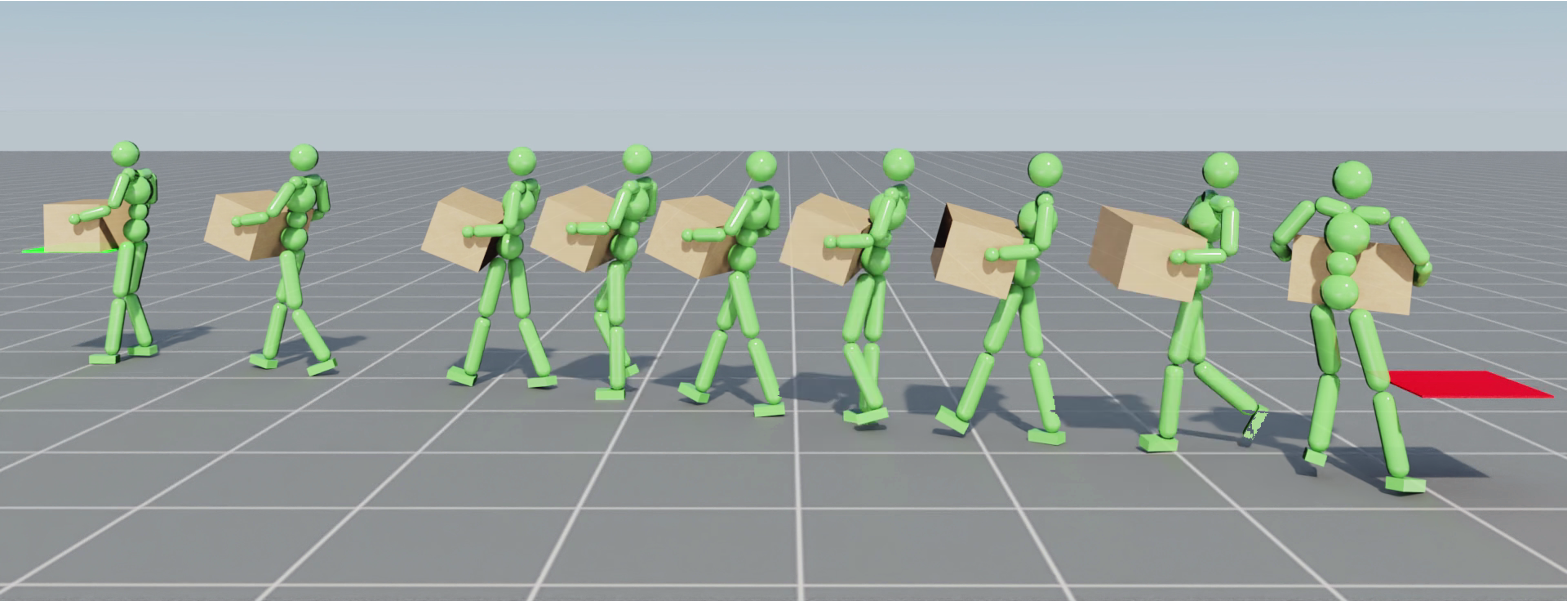}}\\
      \vspace{-0.3cm}
    \caption{Our method successfully executes three challenging scene-interaction tasks in a life-like manner. }
        \label{fig:all_tasks}
\end{figure*}

\section{Results}
In this section, we show results of our method on different scene-interaction tasks. In Fig.~\ref{fig:all_tasks} we show examples of our character executing sit, lie down, and carry tasks. In each task the character is initialized far from the object with a random orientation. The character first approaches the object, using locomotion skills like walking and running, and then seamlessly transitions to task-specific behavior, such as sitting, lying down, or picking up the object. The character is able to smoothly transition from idling to walking, and from walking to the various task-specific behaviors.
For the carry task, note that the object is not attached to the character’s hand, and is instead simulated as a rigid body and moved by forces applied by the character.

\begin{figure*}
	\centering	
 	\includegraphics[trim=000mm 000mm 000mm 017mm, clip=true, width=\linewidth]{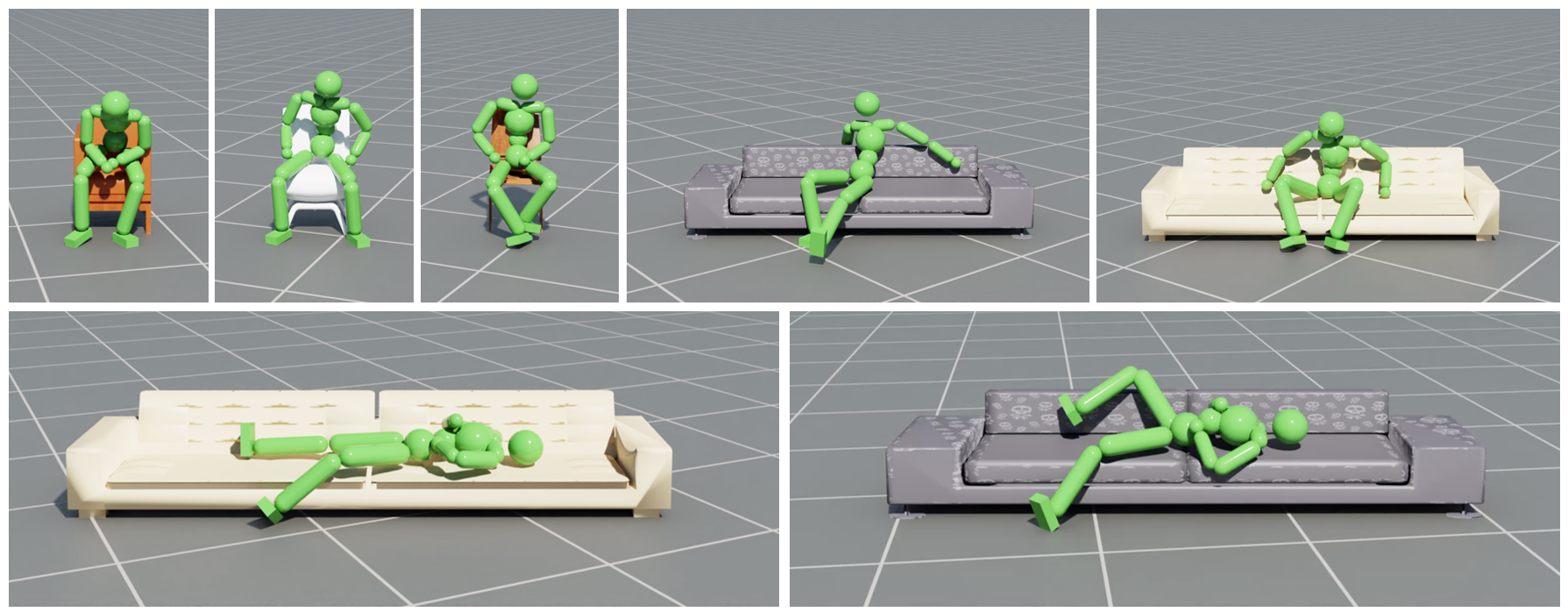}	
	\caption{
	Our method successfully sits and lies down on a wide range of objects and is able to adapt the character's behaviors to new objects.
	}
	\label{fig:geometries_1}
\end{figure*}

\begin{figure}
	\centering	
	\includegraphics[trim=275mm 000mm 000mm 000mm, clip=true, width=\linewidth]{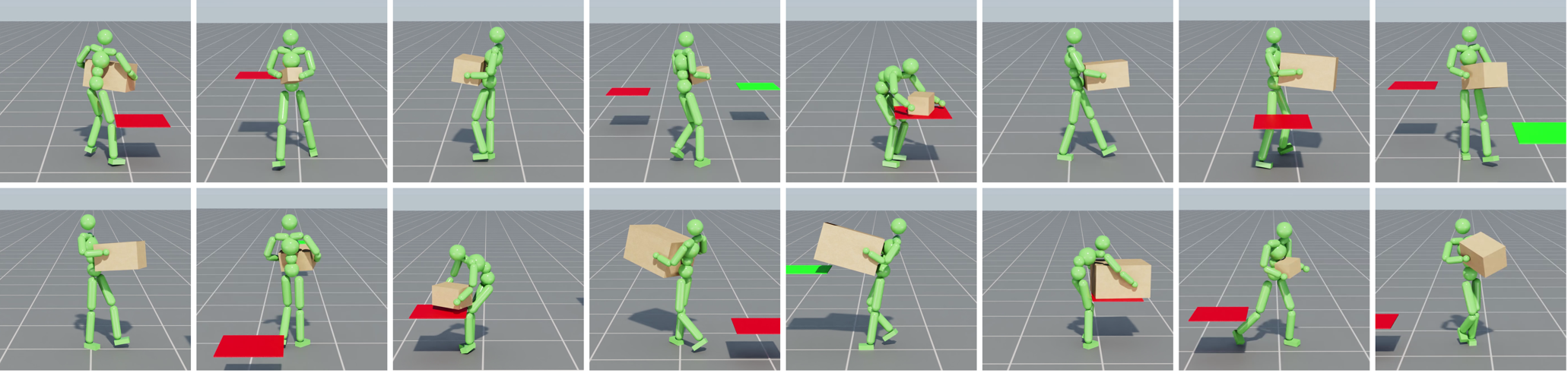}	
	\caption{
	From a human demonstration of carrying a single box, our method generalizes to carrying boxes of different sizes.
	}
	\label{fig:geometries_2}
\end{figure}

\begin{figure}
	\centering	
	\includegraphics[width=\linewidth]{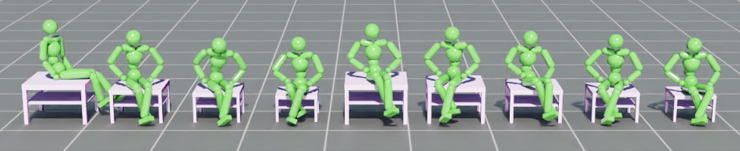}	
	\caption{
	Our policy is able to adapt to different sized objects.
	}
	\label{fig:scales}
\end{figure}

From human demonstrations of interacting with seven objects only, we teach our policy to sit and lie down on $\sim350$ training objects.
We demonstrate the generalization capabilities of our model by testing on objects that were not seen during training as shown in Fig.~\ref{fig:geometries_1}. Our method successfully sits and lies down on a wide range of objects and is able to adapt the character's behaviors accordingly to a given object. The character jumps to sit on a high chair, leans back on a sofa, and puts its arms on the armrests of a chair when present. We used $\sim 350$ training objects and tested on $21$ new objects.
Similarly, our policy learns to carry boxes of different sizes as shown in Fig.~\ref{fig:geometries_2}. We tested our policy on box sizes sampled uniformly between $25 \times 17.5 \times 15$cm and $75 \times 52.5 \times 45$cm.
Our method generalizes beyond what is shown in the original human demonstrations. For example, the character can carry very small boxes as shown in Fig.~\ref{fig:geometries_2}, although no such objects were depicted in the human demonstration dataset. We further test our policy on different scales of the same object as shown in Fig.~\ref{fig:scales}. We observe that the policy learns to adapt to the different sized objects in order to successfully sit or lie down on the support surface. More examples are available in the supplementary \video.

\begin{figure}
	\centering	
	\includegraphics[trim=000mm 7mm 000mm 17mm, clip=true, width=\linewidth]{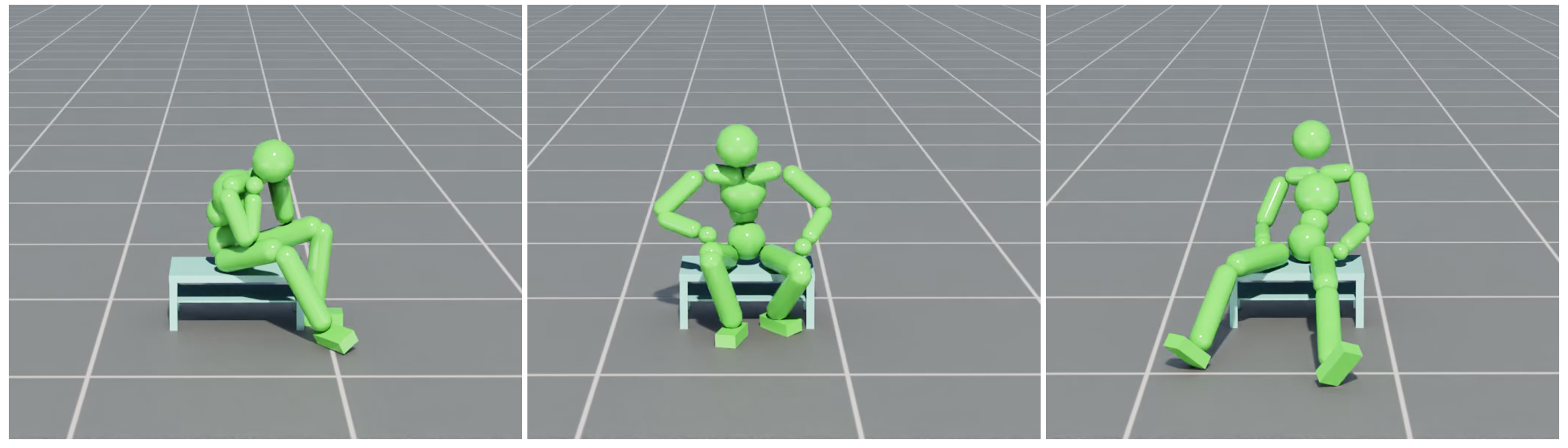}	
	\caption{
	Different styles of sitting on the same object.
	}
	\label{fig:styles}
\end{figure}

Humans have the ability to interact with the same object in a myriad of different styles. As shown in Fig.~\ref{fig:styles}, our character also demonstrates diversity in its interactions with a given object. The character exhibits different styles while sitting, including regular sitting, 
leaning backwards, or sitting with different arms movements.
\subsection{Evaluation}
We quantitatively evaluate our method by measuring the success rate for each task. Table~\ref{tab:success_rate} summarizes the performance statistics on the various tasks. Success rate records the percentage of trials where the character successfully completes the task objectives.
We consider sitting to be successful if the character's hip is within $20$ cm of the target location. Similarly, we declare lying down to be successful if the hip and the head of the character are both within $30$ cm from a target location. 
The carry task is successful if the box is within $20$ cm of the target location. All tasks are considered unsuccessful if their success criterion is not met within $20$ seconds. We evaluate the sit and lie down tasks on $16$ and $5$ unseen objects respectively. To increase the variability between the objects, we randomly scale the objects at each trial with a scale factor between $0.8$ and $1.2$. For the carry task, we randomly scale the original box shown in the human demonstration by a scale factor between $0.5$ and $1.5$ in each trial. The default box has a size of $50 \times 35 \times 30$ cm. The character is randomly initialized anywhere between $1$ m and $10$ m away from the object and with a random orientation. In addition to the success rate, we also measure the average execution time and precision for all successful trials. Execution time is the average time until the character succeeds in executing the task, according to the success definitions above. Precision is the average distance between the hip, head, box and their target locations for sit, lie down, and carry respectively. All metrics are evaluated over $4096$ trials per task. Similarly, we evaluate our carry policy, which is trained to carry boxes of the same size but different weights, in Table~\ref{tab:success_rate} using the same metrics. Please refer to the supplementary material for more details. Despite the diversity of test objects and configurations, our policies succeed in executing all task with a higher than $90\%$ success rate.
\begin{table}[]

\caption{Success rate, average execution time, and average precision for all tasks. All metrics are averaged over $4096$ trails per task.}
\vspace{-0.2cm}
\label{tab:success_rate}
\begin{tabular}{|c|c|c|c|}
\hline
\textbf{Task}   & \textbf{\begin{tabular}[c]{@{}c@{}}Success Rate\\ (\%)\end{tabular}} & \textbf{\begin{tabular}[c]{@{}c@{}}Execution Time\\  (Seconds)\end{tabular}} & \textbf{\begin{tabular}[c]{@{}c@{}}Precision\\  (cm)\end{tabular}} \\ \hline
Sit             & 90.4                                                                 & 5.0                                                                          & 6.7                                                                \\ \hline
Lie down        & 90.2                                                                 & 6.3                                                                          & 13.4                                                               \\ \hline
Carry           & 94.3                                                                 & 9.1                                                                         & 8.3                                                                \\ \hline
Carry (weights) & 97.2                                                                 & 8.7                                                                          & 10.3                                                               \\ \hline
\end{tabular}

\end{table}

\begin{figure}
     \centering
     \subfigure[Sit]{\includegraphics[trim=000mm 025mm 000mm 025mm, clip=true, width=\linewidth]{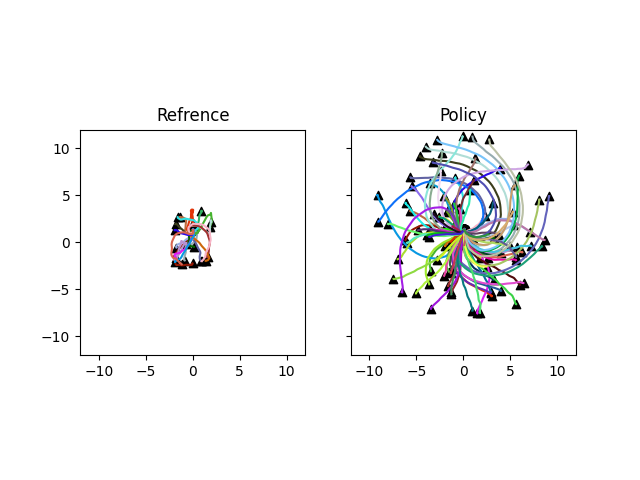}} \\
     \vspace{-0.3cm}
     \subfigure[Lie down]{\includegraphics[trim=000mm 025mm 000mm 025mm, clip=true, width=\linewidth]{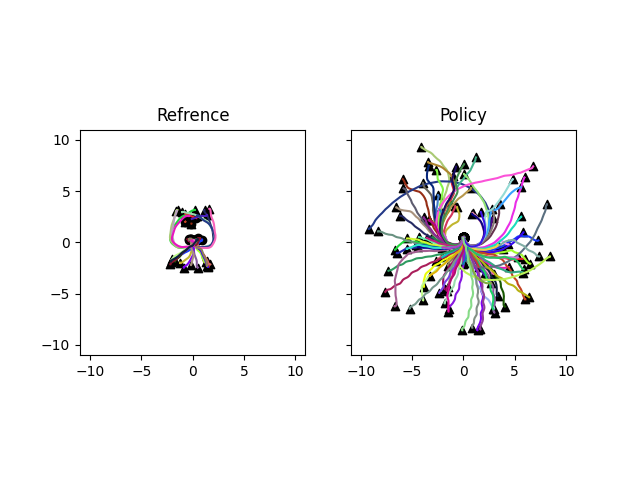}}\\
     \vspace{-0.3cm}
      \subfigure[Carry ]{\includegraphics[trim=000mm 025mm 000mm 025mm, clip=true, width=\linewidth]{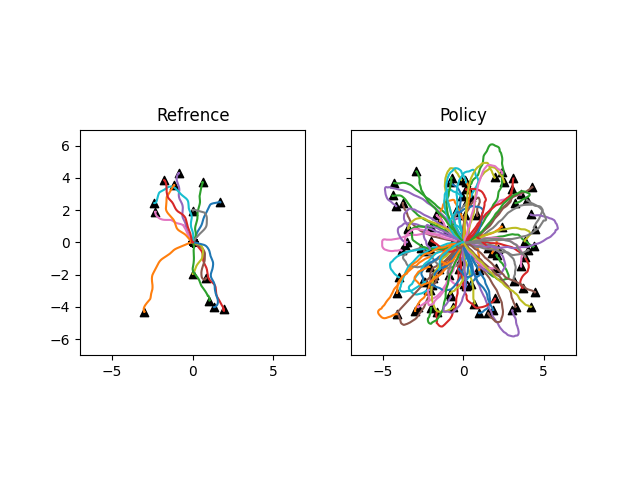}}\\
     \vspace{-0.2cm}

        \caption{Reference motion trajectories and the trajectories generated by our policies when initialized randomly. Triangles indicate starting positions and the target position is indicated with a circle.
         From limited reference clips covering limited  configurations, our policy learns to successfully execute the actions in a wide range of configurations. }
        \label{fig:traj_random}
\end{figure}

Moreover, the character is able to generalize beyond the limited reference clips and succeeds in executing the tasks from initial configurations not shown in the reference motion as can be seen in Fig.~\ref{fig:traj_random}. In the reference clips, the character starts up to three meters away from the object, nonetheless the character learns to execute the tasks even when initialized up to ten meters away from the object. This is partly due to the scene randomization approach used during training as described in Sec.~\ref{sec:training}.

Next we study the robustness of our policy to external perturbations. We pelt the character with $20$ projectiles of weight $1.2$ kg at random time steps of the trial. 
We found that our policy is very robust to these perturbations, and is able to recover and resume the task upon being hit by a projectile. Examples of these recovery behaviors are shown in the supplementary \video.
We also randomly move the object during the execution of a task (e.g.~move the chair away as the character is about to sit). The supplementary \video shows the robustness of the policy to such sudden changes to the environment.
Our policies maintain a high success rate under these physical perturbations for all three tasks, as reported in Table~\ref{tab:success_rate_perturb}
\begin{table}[]
\caption{Success rate under physical perturbations.}
\vspace{-0.2cm}
\begin{tabular}{|c|c|c|}
\hline
\textbf{Task}     & \textbf{Success Rate (\%)} \\ \hline
Sit      &   $87.5$      \\ \hline
Lie down &    $82.0$    \\ \hline
Carry    &     $89.4$   \\ \hline
\end{tabular}
\label{tab:success_rate_perturb}
\end{table}.

\subsection{Comparisons}
There have only been a few previous attempts in the area of synthesizing character-scene interactions. We compare our physics-based model to NSM~\cite{nsm_2019} and SAMP~\cite{hassan_samp_2021}, which are both kinematic models. We also compare to \citet{chao2019learning_to_sit}, which is a hierarchical-based physical approach. 
All three methods are trained on the sitting task. Kinematic models (NSM and SAMP) tend to produce non-physical behaviors, such as foot-skating/floating and object penetrations. Some examples are shown in the supplementary \video. Since, kinematic models learn from human demonstration only, without interaction with the environment,
these models can struggle to generalize to new scenarios. The work of \citet{chao2019learning_to_sit} synthesizes motions using a physics simulation, however it often fails to sit on the target object.
Most of time the character falls when approaching the object.

A quantitative comparison to previous methods is available in Table~\ref{tab:metrics}. A trial is considered successful, only if character does not penetrate the object while approaching it.
None of the baselines are capable of consistently completing the full carry task. NSM~\cite{nsm_2019} trains a character to walk towards a box and lift it up.
However, the character needs to be manually controlled to carry the box to a destination. Our policy, on the other hand, enables the character to autonomously walk towards a box, lift the box, and \textit{carry} it to the destination. 
We use the pre-trained open-source models of NSM~\cite{nsm_2019}, and SAMP~\cite{hassan_samp_2021}, and evaluate them on the same test objects as our method. Note that our method and SAMP are trained on the same dataset. Retraining NSM is infeasible due to the missing phase labels. For \citet{chao2019learning_to_sit}, we report the numbers provided in the paper. Table~\ref{tab:metrics} shows that our method significantly outperforms these prior systems on the sit and lie down tasks.

\section{Discussion}
Throughout our experiments, we train a separate policy for each task. Multi-task RL remains a difficult and open problem~\cite{ruder2017overview} and should be investigated  in future work.
Unlike previous attempts to synthesize carry motions \cite{coros_2010, peng_2019,mordatch_2012}, our box is not welded to the character’s hand. The box is simulated as a rigid object and is moved by forces applied by the character. 
In a few cases, the character approaches the object but fails to complete the task successfully within the duration of an episode. For example, the character might stand next to the object until the end of the episode. In other cases, the character might not reach the target object in time because it follows a suboptimal path; some examples are shown in Fig.~\ref{fig:traj_random}. 
We focus on environments of one objects only. Nonetheless, our state representation could be augmented to contain other objects. In addition, it would be exciting to explore adding virtual eyes to our character. This would allow for interaction with more complex scenes.
We show quantitatively and qualitatively that our randomization approach enables the character to interact with a wide range of test objects. These objects are not used during training and are randomly selected from ShapeNet. We also show that our method can adapt to different object sizes (Fig.~\ref{fig:geometries_2} and Fig.~\ref{fig:scales}) and weights (Table.~\ref{tab:success_rate}). Nonetheless, if the test size or weight is far from the training distribution, we expect the success rate to drop.
We focus on generalization to different objects, future work should explore generalization to different skills, such as jumping.

\begin{table}[]
\caption{Performance comparision to NSM~\cite{nsm_2019}, SAMP~\cite{hassan_samp_2021}, \citet{chao2019learning_to_sit}}
\label{tab:metrics}
\resizebox{\linewidth}{!}
{
\begin{tabular}{|c|cccc|cc|}
\hline
\multirow{2}{*}{\textbf{Metric}} & \multicolumn{4}{c|}{\textbf{Sit}}                                                                          & \multicolumn{2}{c|}{\textbf{Lie down}}            \\ \cline{2-7} 
                                 & \multicolumn{1}{c|}{NSM}  & \multicolumn{1}{c|}{SAMP}          & \multicolumn{1}{c|}{Chao \etal} & Ours          & \multicolumn{1}{c|}{SAMP}          & Ours         \\ \hline
Success Rate(\%)                 & \multicolumn{1}{c|}{75.0} & \multicolumn{1}{c|}{75.0}          & \multicolumn{1}{c|}{17}   & \textbf{93.7} & \multicolumn{1}{c|}{50}            & \textbf{80}  \\ \hline
Execution Time(seconds)          & \multicolumn{1}{c|}{7.5}  & \multicolumn{1}{c|}{7.2}           & \multicolumn{1}{c|}{-}    & \textbf{3.7}  & \multicolumn{1}{c|}{9.5}           & \textbf{6.9} \\ \hline
Precision (meters)               & \multicolumn{1}{c|}{0.19} & \multicolumn{1}{c|}{\textbf{0.06}} & \multicolumn{1}{c|}{-}    & 0.09          & \multicolumn{1}{c|}{\textbf{0.05}} & 0.3          \\ \hline
\end{tabular}
}
\end{table}

\section{Conclusion}
We presented a method that realistically synthesizes physical and realistic character-scene interaction. We introduced a scene-conditioned policy and discriminator that take into account a character's movements in the context of objects in the environment. We applied our method to three challenging scene interaction tasks: sit, lie down, and carry. Our method learns when and where to transition from one behavior to another to execute the desired task. We introduced an efficient randomization approach for the training objects, their placements, sizes, and physical properties. This randomization approach allows our policies to generalize to a wide range of objects and scenarios not shown in the human demonstration. We showed that our policies are robust to different physical perturbations and sudden changes in the environment. 
We qualitatively and quantitatively showed that our method significantly outperforms previous systems. We hope our system provides a step towards creating more capable physically simulated characters that can interact with complex environments in a more intelligent and life-like manner. 

\noindent
\textbf{Disclosure:}
The work was done while Mohamed Hassan was an intern at Nvidia. MJB has received research funds from Adobe, Intel, Nvidia, Facebook, and Amazon. While MJB is a part-time employee of Amazon, his research was performed solely at, and funded solely by, Max Planck. MJB has financial interests in Amazon, Datagen Technologies, and Meshcapade GmbH.

\bibliographystyle{ACM-Reference-Format}
\bibliography{00_bib}

\renewcommand{\thefigure}{S.\arabic{figure}}
\setcounter{figure}{0}
\renewcommand{\thetable}{S.\arabic{table}}
\setcounter{table}{0}

\clearpage
\begin{appendices}

\section{Tasks}\label{sec:tasks}
Our aim is to train simulated character to solve character-scene interaction tasks. To demonstrate the effectiveness of our method; we choose three challenging interactive tasks: sit, lie down, and carry. The style reward $r^S$ is the same for all tasks as defined. The task reward $r^G$ is task-specific as detailed in the following subsections.

\subsection{Sit}
The objective of this task is for the character to move to a target object and to sit on it.The object is initialized at a random orientation anywhere between one and ten meters away from the character. 

The goal $\rvg_t \in \mathbb{R}^3$ is the object bounding box.
The task reward is defined as :
\begin{equation}
    r^G_t = 
    \begin{cases}
        0.7 \ r^\mathrm{near}_t + 0.3 \ r^\mathrm{far}_t,  & ||\rvx^* - \rvx_t^\mathrm{root}|| > 0.5m \\
        0.7 \ r^\mathrm{near}_t + 0.3,  & \text{otherwise}\\
        
    \end{cases}
    \label{eqn:rewardSit}
\end{equation}
where $\rvx^\mathrm{root}$ is the position of the character's root, $\rvx^*$ is the object position, $r^\mathrm{far}$ encourages the character to walk towards the object, while $r^\mathrm{near}$ encourages the character to sit on the object once it is close by. 
$r^\mathrm{far}$ is specified according to:
\begin{align}
    r^\mathrm{far}_t & = 0.5 \ \mathrm{exp}\left(-0.5 ||\rvx^* - \rvx^\mathrm{root}_t||^2 \right) \nonumber \\ 
    & + 0.5 \ \mathrm{exp}\left( -2.0 ||\ v^* - \rvd^*_t \cdot \dot{\rvx}_t^\mathrm{root} ||^2 \right)
\label{eqn:rewardSitFar}
\end{align}
where  $\dot{\rvx}_t^\mathrm{root}$ is the linear velocity of the character's root, $\rvd^*$ is a horizontal unit vector pointing from the root $\rvx_t^\mathrm{root}$ to the object's location $\rvx^*$, and $v^*=1.5 m/s$ is the target speed at which the character should walk. Once the character is close to the object, $r^\mathrm{near}$ is used to encourage the character to sit on the object:
\begin{equation}
    r^\mathrm{near}_t =  \mathrm{exp}\left(-10.0 ||{\rvx^\mathrm{root}}^* - \rvx_t^\mathrm{root}||^2 \right),
    \label{eqn:rewardSitNear}
\end{equation}
 with ${\rvx^\mathrm{root}}^*$ denoting the target sitting position on the object where the character's hip should be placed.

\subsection{Lie down}
 The objective of the lie down task is for the character to walk towards an object and then lie down on it. The goal $\rvg_t$ and the task reward $r^G_t$ are the same as for the sitting task (see Eq.~\ref{eqn:rewardSit}). $r^\mathrm{far}$ is defined as in Eq.~\ref{eqn:rewardSitFar}, and $r^\mathrm{near}_t = $
 \begin{equation}
   \mathrm{exp}\left(-10.0 ||{\rvx^\mathrm{root}}^* - \rvx_t^\mathrm{root}||^2  -10.0 ||{h^\mathrm{head}}^* - h_t^\mathrm{head}||^2 \right)
    \label{eqn:rewardLiedownNear}
\end{equation}
where $h^\mathrm{head}$ is the height of the character's head, and ${\rvx^\mathrm{head}}^*$ is the target head height.

\subsection{Carry}
The objective of the carry task is for the character to pick up a box and carry it to a destination. The goal is specified according to:
\begin{equation}
    \rvg_t=( \tilde{\rvx}'_t, b_h, b_w, b_d),
    \label{eqn:carryGoal}
\end{equation}
where  $\tilde{\rvx}'_t$ is the target position on which the box should be placed; $\tilde{\rvx}'_t$ is represented in the character's local coordinate frame. $b_h, b_w, b_d$ are the box height, width, and depth respectively.
The task reward is specified according to:
\begin{equation}
    r^G_t = r^\mathrm{walk}_t + r^\mathrm{carry}_t,
    \label{eqn:rewardCarry}
\end{equation}
where $r^\mathrm{walk}$ encourages the character to walk towards the box and stay close to it. More specifically, it encourages the character to move its root $\rvx^\mathrm{root}_t$ towards the position of the box $\rvx^*_t$ at a target speed $\rvd^*$:
\begin{equation}
    r^\mathrm{walk}_t = 
    \begin{cases}
        0.1 \ \mathrm{exp}\left(-0.5 ||\rvx^*_t - \rvx^\mathrm{root}_t||^2 \right) +  \nonumber \\
     0.1 \ \mathrm{exp}\left( -2.0 ||\ v^* - \rvd^*_t \cdot \dot{\rvx}_t^\mathrm{root} ||^2 \right), & ||\rvx^*_t - \rvx_t^\mathrm{root}|| > 0.5m \\ 
     0.2, & \text{otherwise} 
    \end{cases} .
\end{equation}
$r^\mathrm{carry}$ encourages the character to carry the box to a target position $\rvx'_t$:
\begin{equation}
    r^\mathrm{carry}_t = 
    \begin{cases}
        r^\mathrm{carry-far}_t + r^\mathrm{carry-near}_t , & ||\rvx'_t - \rvx^*_t|| > 0.5m \\ 
     0.2 + r^\mathrm{carry-near}_t, & \text{otherwise} 
    \end{cases} .
\end{equation}
$r^\mathrm{carry-far}_t$ is defined as: 
\begin{align}
    r^\mathrm{carry-far}_t & = 0.2 \ \mathrm{exp}\left(-0.5 ||\rvx'_t - \rvx^*_t||^2 \right) \nonumber \\ 
    & + 0.2 \ \mathrm{exp}\left( -2.0 ||\ v' - \rvd'_t \cdot \dot{\rvx}_t^\mathrm{box} ||^2 \right) \nonumber \\
    & + 0.1 \ \mathrm{exp} \left( -10.0 ||{h_t^\mathrm{hand}} - h_t^\mathrm{box}||^2 \right)
    .
\end{align}
Where $\rvd'_t$ is a horizontal unit vector pointing from the box location to the target location, $\dot{\rvx}_t^\mathrm{box}$ is the velocity of the box, $v'=1.5 m/s$ is the target speed. $h_t^\mathrm{hand}$ and $h_t^\mathrm{box}$ are the height of the character's hand and box height respectively. 
Once the box is close to the target, $r^\mathrm{carry-near}_t$ encourages the character to place the box precisely on the target platform,
\begin{align}
    r^\mathrm{carry-near}_t & = 0.2 \ \mathrm{exp}\left(-10.0 ||\rvx'_t - \rvx^*_t||^2 \right).
\end{align}

\section{Results}
\begin{figure}
	\centering	
	\includegraphics[width=\linewidth]{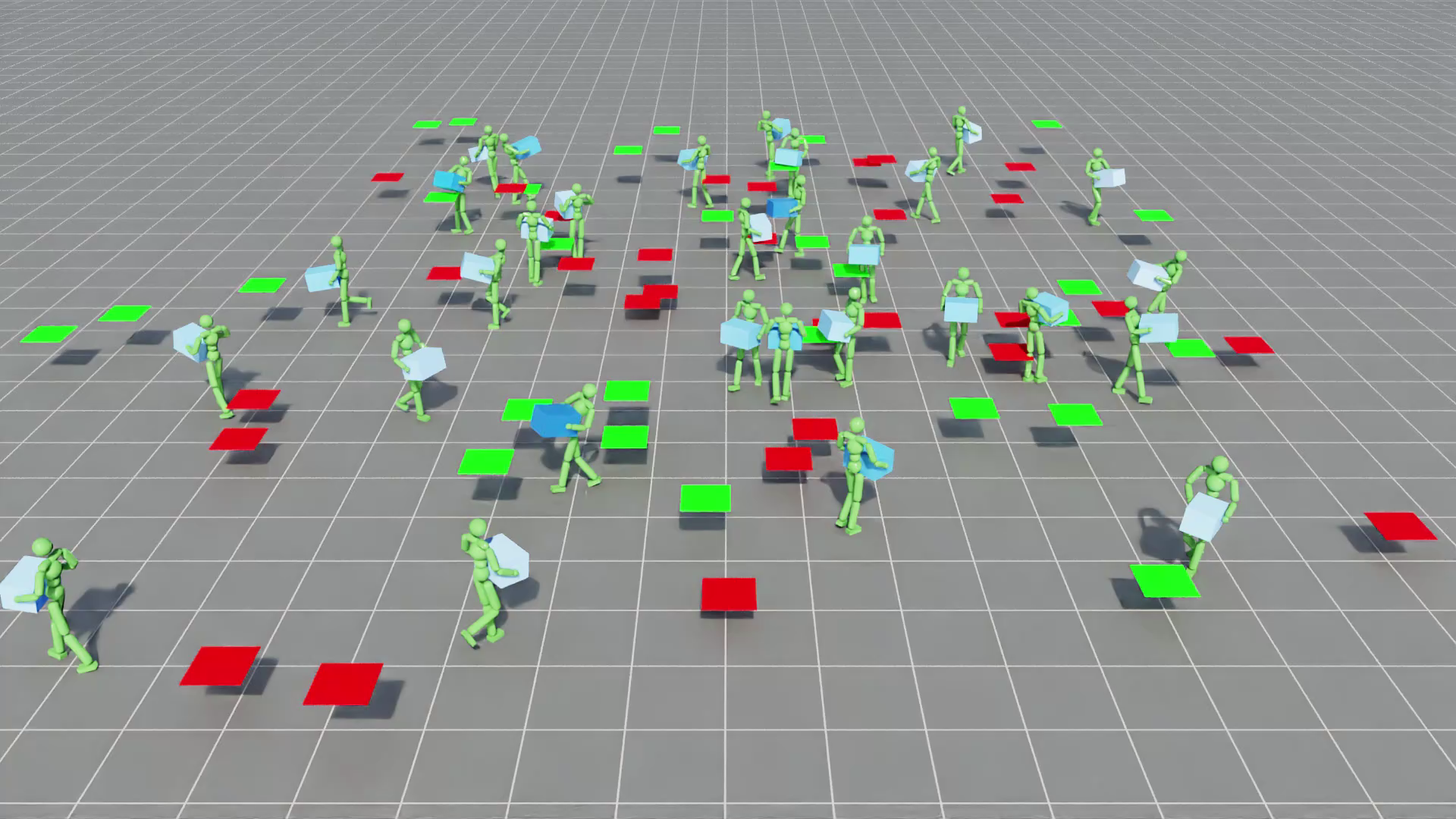}	
	\caption{
	Carrying boxes of different weights. Darker colors indicate heavier weights.
	}
	\label{fig:weights}
\end{figure}

We show how our policy deals with objects of different physical properties. We train a policy to carry boxes of the same size but different weights. The weights are sampled uniformly between $5$ kg and $26$ kg. For this experiment, we augment the goal $\rvg_t$ with the box density. Some examples are shown in Fig.~\ref{fig:weights} where heavier boxes are indicated with darker colors. The character discovers how to deal with the different weights from a human demonstration of a single box.

Note that our reference data only contains demonstrations of carrying a single box in a particular style. Thus, it is expected that the character will stay close to this carrying style, even for boxes of different weights. The purpose of this experiment is to show carrying objects with weights that are different from the reference data. To learn different carrying styles, we would need such styles to be part of our dataset.

\begin{figure}
     \centering
     \subfigure[Sit]{\label{fig:traj_ref_sit}\includegraphics[trim=000mm 025mm 000mm 025mm, clip=true, width=\linewidth]{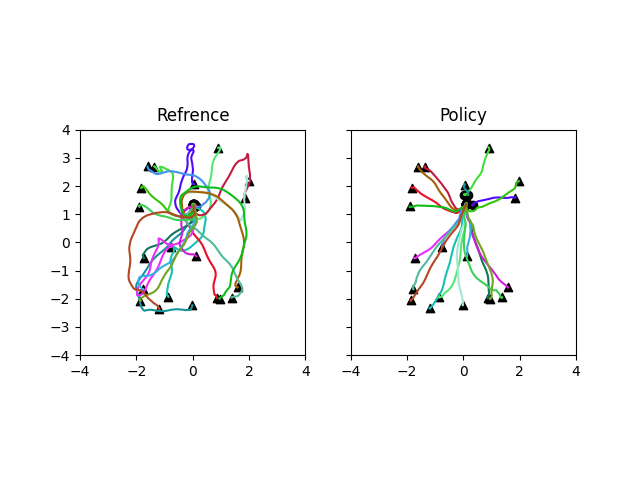}}
     \subfigure[Lie down]{\includegraphics[trim=000mm 025mm 000mm 025mm, clip=true, width=\linewidth]{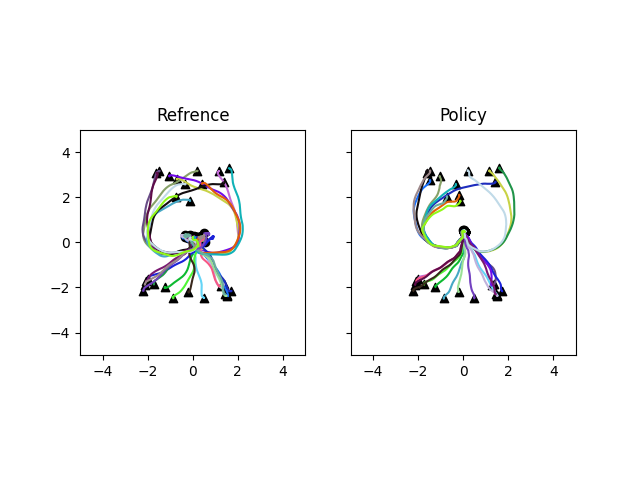}}
      \subfigure[Carry ]{\includegraphics[trim=000mm 025mm 000mm 025mm, clip=true, width=\linewidth]{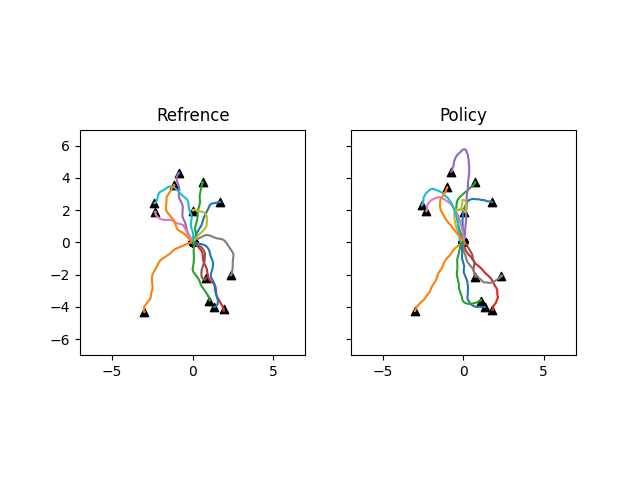}}

        \caption{Reference motion trajectories and the trajectories generated by our policies when initialized with the first frame of the reference motion. Triangles indicate starting positions and the target position is indicated with a circle.
         Although the reference clips do not always follow the shortest trajectory to the object, our policy often does. }
        \label{fig:traj_ref}
\end{figure}

We illustrate the plausibility of the full motion trajectories generated by our policy in Fig.~\ref{fig:traj_ref}. We initialize our policy with the first frames of the reference motion clips. We then plot the full trajectories followed by our policy alongside the reference trajectories from the reference motion clips. For the sit and lie down tasks, we plot the character trajectory. For the carry task, we plot the box trajectory. Although the reference clips do not always follow the shortest trajectory to the object, our policy often does as can be seen in Fig.~\ref{fig:traj_ref_sit}. 

Throughout our experiments, we include the bounding box of the object in our goal $\rvg_t$ as explained in Sec.~\ref{sec:tasks}. To evaluate the importance of the bounding box, we retrain our policies without this information and evaluate the policies in Table~\ref{tab:bb_ablation}. We observe that the bounding box is vital especially for dynamic tasks like carry. Without this information the character fails to pick up the box from the platform. In general, excluding the bounding box information decrease the success rate for all three tasks.
\begin{table}[]

\caption{Bounding box ablation. Success rate, average execution time, and average precision for all tasks. }
\label{tab:bb_ablation}
\begin{tabular}{|c|c|c|c|}
\hline
\textbf{Task}   & \textbf{\begin{tabular}[c]{@{}c@{}}Success Rate\\ (\%)\end{tabular}} & \textbf{\begin{tabular}[c]{@{}c@{}}Execution Time\\  (Seconds)\end{tabular}} & \textbf{\begin{tabular}[c]{@{}c@{}}Precision\\  (cm)\end{tabular}} \\ \hline
Sit             & 88.6                                                                 & 5.3                                                                          & 6.6                                                                \\ \hline
Lie down        & 81.9                                                                 & 6.2                                                                          & 14.7                                                               \\ \hline
Carry           & 0.0                                                                 & -                                                                         & -                                                                \\ \hline

\end{tabular}

\end{table}

\section{Comparisons}
NSM~\cite{nsm_2019} and SAMP~\cite{hassan_samp_2021} have two modes of operation. Manual mode, where a user controls the character, and a goal-driven mode. In the goal-driven mode, the user only specifies the desired action(e.g. sit) and the target object. We are only comparing against the goal-driven mode of NSM and SAMP. Our method, NSM, and SAMP all receive the same input from the user which is the desired action and the target object. The full motion is generated by the algorithms and there is no manual nor random control.

\end{appendices}

\end{document}